**This is the accepted manuscript (postprint) of the following article:**
E. Salahinejad, A. Muralidharan, F.A. Sayahpour, M. Kianpour, M. Akbarian, D. Vashaee, L. Tayebi, *Enhanced vascularity in gelatin scaffolds via copper-doped magnesium–calcium silicates incorporation: In-vitro and ex-ovo insights*, Ceramics International, 50 (2024) 39889-39897.
https://doi.org/10.1016/j.ceramint.2024.07.369

# Enhanced Vascularity in Gelatin Scaffolds via Copper-Doped Magnesium-Calcium Silicates Incorporation: *In-Vitro* and *Ex-Ovo* Insights

Erfan Salahinejad [a, b], Avaneesh Muralidharan [a], Forough Azam Sayahpour [a], Maryam Kianpour [a], Mohsen Akbarian [a], Daryoosh Vashaee [c,d], Lobat Tayebi [a,*]

[a] *Marquette University School of Dentistry, Milwaukee, WI 53233, USA*

[b] *Faculty of Materials Science and Engineering, K. N. Toosi University of Technology, Tehran, Iran*

[c] *Department of Electrical and Computer Engineering, North Carolina State University, Raleigh, NC 27606, USA*

[d] *Department of Materials Science and Engineering, North Carolina State University, Raleigh, NC 27606, USA*

## Abstract

Addressing a critical challenge in current tissue-engineering practices, this study aims to enhance vascularization in 3D porous scaffolds by incorporating bioceramics laden with pro-angiogenic ions. Specifically, freeze-dried gelatin-based scaffolds were infused with sol-gel-derived powders of Cu-doped akermanite ($Ca_2MgSi2O_7$) and bredigite ($Ca_7MgSi_4O_{16}$) at various concentrations (10, 20, and 30 wt%). The scaffolds were initially characterized for their structural integrity, biodegradability, swelling behavior, impact on physiological pH, and cytocompatibility with human umbilical vein endothelial cells (HUVECs). The silicate incorporation effectiveness in promoting vascularity was then assessed through HUVEC attachment, capillary tube formation, and *ex-ovo* chick embryo chorioallantoic membrane assays. The findings revealed significant improvements in both *in-vitro* and *ex-ovo* vascularity of the gelatin scaffolds upon the addition of Cu-doped akermanite. The most effective concentrations were determined to be 10 and 20%, which led to notable HUVEC metabolic activity, a well-spread morphology with extensive peripheral filopodia and lamellipodia at 10% and a cobblestone phenotype indicative of *in-vivo* endothelium at 20% during cell attachment, the formation of complex networks of tubular structures, and robust vascularization in chick embryo development. Moving forward, the incorporation of Cu-doped akermanite into tissue-engineering scaffolds shows great potential for addressing the limitations of vascularization, especially for critical-sized bone defects, by facilitating the controlled release of pro-angiogenic and pro-osteogenic ions.

[*] Corresponding Author: LT <lobat.tayebi@marquette.edu>

## 1. Introduction

Tissue engineering plays a critical role in modern medicine due to its ability to regenerate tissues and organs, providing personalized solutions and diminishing dependence on donors. Gelatin as an enzymatically-degradable biopolymer is often used in scaffold-based tissue-engineering studies, mostly due to its biocompatibility, biodegradability, hydrogel-forming ability, and cost-effectiveness [1-3]. However, this collagen derivative does not solely suffice for effective tissue regeneration, especially to treat critical-sized defects. Typical drawbacks of gelatin include extreme biodegradability, inadequate mechanical properties, and limited vascularity. Although methods such as cross-linking, blending with other polymers, and incorporating bioceramics can effectively address the biodegradability and mechanical properties of gelatin, the challenge of enhancing vascularity remains inadequately addressed.

The significance of robust vascularization in tissue regeneration is underscored by its ability to ensure the delivery of essential nutrients, oxygen, and signaling molecules, facilitates waste removal, guides cell migration, supports immune responses, and helps maintain an optimal physiological environment for effective tissue repair and regeneration [4-6].

There are several natural factors responsible for triggering vascularization in the human body, such as vascular endothelial growth factor (VEGF) [7, 8], basic fibroblast growth factor (bFGF) [9-11], and hypoxia-inducible factor-1-alpha [12, 13]. Despite the powerful regulatory role of these factors in vascularization, concerns including uncontrolled vascularization, safety issues, delivery challenges, and cost have contributed to researchers to explore alternative approaches for medical intervention. Several plant-derived compounds (such as β-Sitosterol [14] and Calycosin [15]) and some bioactive ions (including boron [16], cobalt [17], copper [18], zinc [19], magnesium [20], strontium [21], and silicon [22]) are illustrative agents that can more safely promote vascularization. Notably, the incorporation of proper bioceramics that release these pro-angiogenic ions can enhance not only vascularity, but also mechanical strength in gelatin-based constructs.

Ca-Mg silicates have accrued significant interest, especially for bone tissue-engineering applications owing to considerable mechanical behaviors, biodegradability, biocompatibility, bioactivity, and osteogenic ability [23-25], including in *in-vivo* studies [26-29]. Their potential to stimulate angiogenesis has been also pointed out in different studies, especially for akermanite ($Ca_2MgSi_2O_7$) [27, 30-33]. However, their angiogenic power is not sufficiently high to be used as an additive in structures that lack vascularity, while there is a need for it in certain applications. He *et al.* [29] indicated that copper doping in akermanite and bredigite ($Ca_7MgSi_4O_{16}$) coatings significantly enhance the early-stage neovascularization of diopside ($CaMgSi_2O_6$) *in-vivo*. Cu is primarily known for its antimicrobial properties through the release of toxic copper ions that disrupt bacterial cell membranes, enzymes, and DNA with generating oxidative stress [34-36]. The pro-angiogenic role of copper is also considerable. It stimulates endothelial cell proliferation and migration, induces the release of angiogenic growth factors like VEGF, influences metalloproteinase activity for tissue remodeling, modulates cellular signaling pathways, generates reactive oxygen species, and interacts with proteins and enzymes involved in angiogenesis [37-39]. However, there are no reports on the employment of pro-angiogenic Cu-doped Ca-Mg silicates in tissue-engineering scaffolds that lack sufficient vascularity, to our knowledge. Accordingly, this work originally hypothesizes that the incorporation of particulate Cu-doped akermanite and bredigite into gelatin-based constructs can effectively improve vascularity.

## 2. Materials and Methods

### *2.1. Preparation of Bioceramic Powders*

Copper-doped akermanite and bredigite powders were synthesized via a sol-gel route, utilizing calcium nitrate tetrahydrate ($Ca(NO_3)_2 \cdot 4H_2O$, >99%, Thermo Scientific Chemicals, USA), magnesium chloride hexahydrate ($MgCl_2 \cdot 6H_2O$, >99%, VWR Chemicals BDH, USA), copper (II) chloride dihydrate ($CuCl_2 \cdot 2H_2O$, >99%, Thermo Scientific Chemicals, USA), and tetraethyl Orthosilicate ($Si(OC_2H_5)_4$, TEOS, >97%, TCI America, USA). The sol-gel process was adapted from Refs. [30, 40, 41] with minor modifications, with the level of Cu incorporation inspired by Ref. [29]. Briefly, TEOS was first hydrolyzed using nitric acid ($HNO_3$, 1 N, VWR Chemicals BDH, USA) at the molar ratio of TEOS/$H_2O$/$HNO_3$ = 1:8:0.16 for akermanite and = 1:4:0.08 for bredigite under stirring for 30 min. The other precursors were introduced to the hydrolyzed solution with the molar ratio of TEOS/$Mg(NO_3)\cdot 6H_2O$/$Ca(NO_3)\cdot 4H_2O$ = 2:1:2 for akermanite and = 4:1:7 for bredigite, while 5 mol% of $Ca(NO_3)\cdot 4H_2O$ was substituted with the same amount of $CuCl_2\cdot 2H_2O$ and the reactants were stirred at ambient temperature for 5 h. After stirring, the solutions were transferred to a water bath and maintained at 60 °C for 24 h. The obtained gels were frozen at -20 °C for 24 h subsequently subjected to freeze drying (Labconco Freezone 9.8) for 2 h to remove unreacted solvents and water. The dry gels were grounded, sieved to prevent agglomerated particles, and calcined at 1150 °C in an electrical furnace (KSL-1700X-S-MTI Corporation). Finally, the calcined Cu-doped akermanite and bredigite powders were again ground, and sieved, designated as Samples A and B, respectively, in this paper.

### *2.2. Fabrication of Porous Gelatin Scaffolds with Embedded Bioceramics*

Porous gelatin-matrix, silicate-incorporated scaffolds were fabricated by freeze drying. Gelatin Type A from porcine skin (Sigma-Aldrich, USA) was dissolved in deionized water at 65 °C under stirring, followed by the addition of certain amounts of Powders A and B. Seven scaffold groups, labeled C, A10, A20, A30, B10, B20, and B30, were fabricated with ceramic weight percentages of 0, 10, 20, and 30 for A and B, respectively. The suspensions were transferred to a mechanical mixer (EUROSTAR 60 control - IKA) to create gelatinous foams. The foams were poured into small petri dishes, frozen overnight at -20 °C, and freeze dried at 3 Pa, -50 °C to remove all water from the scaffolds. The samples were then crosslinked in an ethanolic solution containing 2% 1-Ethyl-3-(3-dimethylaminopropyl) carbodiimide hydrochloride (EDC, 99%, Oakwood Chemical, USA) and 0.4% N-hydroxysuccinimide (NHS, >98%, Alfa Aesar, USA) at 4 °C for 24 h. Unreacted EDC and NHS were washed out thoroughly using phosphate buffered saline (PBS). The crosslinked scaffolds were finally frozen overnight and freeze-dried for 12 h.

### *2.3. Characterization*

#### *2.3.1. Structure*

The phase characterization of the ceramic powders was conducted using X-ray diffraction (XRD, Bruker AXS D8-Discover, Cu Kα irradiation, λ = 0.15406 nm, step size = 2°, scan rate = 2°/min) and analyzed with PANalytical X'Pert HighScore software. The porosity of the scaffolds was determined through the Archimedes' ethanol immersion method, ASTM C20-00 [42], performed in triplicate. Additionally, the morphology of the porous constructs was analyzed using a scanning electron microscope (SEM, JEOL JSM-6510LV) with an accelerating voltage of 15 kV.

#### *2.3.2. Physiochemical Behaviors*

To evaluate the hydrolytic degradation and swelling characteristics of the scaffolds, along with changes in the pH of the physiological medium to which they were exposed, the scaffolds were cut into disks of 3 mm in diameter and 3 mm in thickness. The specimens were first weighted ($W_i$) using a NewClassic MS Mettler Toledo analytical balance, soaked in 10 ml PBS, and placed in a 37 °C incubator. At certain timeframes, the scaffolds were removed and dried at an oven of 40 °C for 24 h, while PBS was refreshed to continue the experiments. Afterward, the weight of the dried samples was measured as $W_f$, and degradation levels were determined using the following formula with three repetitions:

$$\text{Degradation Level (\%)} = \left(\frac{W_i - W_f}{W_i}\right) \times 100 \tag{1}$$

For swelling characterizations, after measuring the initial weight of the samples, they were submerged in 10 ml PBS at 37 °C, taken out of the solution at certain intervals, gently surface-dried using a filter paper, and then weighed in their wet states ($W_w$), with PBS refreshing. The swelling degree was calculated by the following equation with three repetitions:

$$\text{Swelling Ratio} = \left(\frac{W_w - W_i}{W_i}\right) \tag{2}$$

Also, the pH value of PBS in contact with the scaffolds was measured without refreshing using Fisher Scientific Accumet Basic AB15 pH meter with three repetitions.

#### ***2.3.3. Cell cytocompatibility***

The MTT assay was employed to explore the metabolic activity of human umbilical vein endothelial cells (HUVECs) on the different samples with the pure gelatin sample (C) as the control group. The samples, cut into disks measuring 3 mm in diameter and 3 mm in thickness, were subjected to sterilization through UV exposure for 2 h and treatment in 70% ethanol for 3 h. Subsequently, they were rehydrated overnight in an endothelial cell basal medium (EBM$^{TM}$-2, Lonza, USA) supplemented with an endothelial cell growth medium (EGM$^{TM}$-2 SingleQuots, Lonza, USA, excluding VEGF to highlight the inherent pro-angiogenic potential of the scaffolds) and 1% penicillin-streptomycin. After seeding $3.0\times10^4$ HUVECs on the specimens in 48-well cell culture plates, the samples were incubated at 37 °C under a 5% $CO_2$ atmosphere for periods of 24, 48, and 72 h. The MTT assay was conduct by measuring the absorbance at 570 nm by a BioTek Synergy HTX multi-mode microplate reader. The collected data, acquired through three repetitions, underwent one-way analysis of variance (ANOVA), with statistical significance established at $P < 0.05$.

#### ***2.3.4. Vascularity***

HUVEC attachment, tube formation, and chick embryo assays were conducted to investigate the vascularity of the samples. As a primary indicative measure of vascularization, the attachment of HUVECs to the surfaces was studied morphologically by culturing $1.0\times10^5$ cells on each scaffold for 5 days. Then, the cells were fixed using a sequence of 2.5% glutaraldehyde overnight and then 30, 50, 60, 70, 80, 90, and 100% ethanol, each for 30 min, followed by gold plating and SEM observations at the accelerating voltage of 15 kV.

For the tube formation assay, extractions were collected from the scaffolds incubated in the specific endothelial cell medium for 24 h. Subsequently, $1.5\times10^4$ HUVECs were seeded on the Corning Matrigel Basement Membrane Matrix with the extracted media for 12 h, followed by Calcein AM staining and fluorescence microscopy (Life Technologies EVOS FL Auto).

The *ex-ovo* chick chorioallantoic membrane assay was utilized to further study the influence of the scaffolds on angiogenesis in Barred Plymouth rock hatching eggs sourced from Meyer Hatchery Co, USA. Initially, the eggs were placed in a chicken hatching incubator (Sailnovo) set at 78% humidity and 38 ºC for 3 days with a rotation rate of 45° every two hours. On the third day, following the confirmation of vessel formation in the eggs using a concentrated light, the eggshells were carefully broken, and the whole contents were then transferred into a series of Corning® 100 mm culture dishes. Subsequently, the UV-sterilized hydrogels in pieces of identical size and weight were meticulously placed on chorioallantoic membrane regions. Following a 48-h incubation period, optical photographs were taken to observe the angiogenesis process.

## 3. Results and Discussion

### *3.1. Structural Characterization*

The XRD pattern of the calcined ceramic powders, as shown in Fig. 1a, reveals well-defined, intense peaks indicating the samples' significant crystallinity. This crystallinity is attributed to the elevated-temperature calcination process applied. The peak analysis revealed that Samples A and B consist of akermanite and bredigite mono-phases, respectively. The absence of detectable secondary phases suggests that the ion release kinetics from the samples, upon exposure to physiological media, will be consistent with the established degradation rates of these phases [43-45]. Utilizing the modified Scherrer Equation [46, 47], the average crystallite sizes for akermanite and bredigite were determined to be approximately 30 and 35 nm, respectively. This is indicative of bottom-up nanocrystallization from sol-gel-derived amorphous phases. Peak shifts of approximately 0.2° toward higher Bragg angles, compared to pure akermanite ($Ca_2MgSi_2O_7$) and bredigite ($Ca_7MgSi_4O_{16}$) with ICDD reference codes 35-0592 and 36-0399, respectively, were observed. These shifts are attributed to the introduction of Cu doping into the phases, leading to a reduction in their lattice parameters due to the smaller ionic radius of $Cu^{2+}$ compared to $Ca^{2+}$.

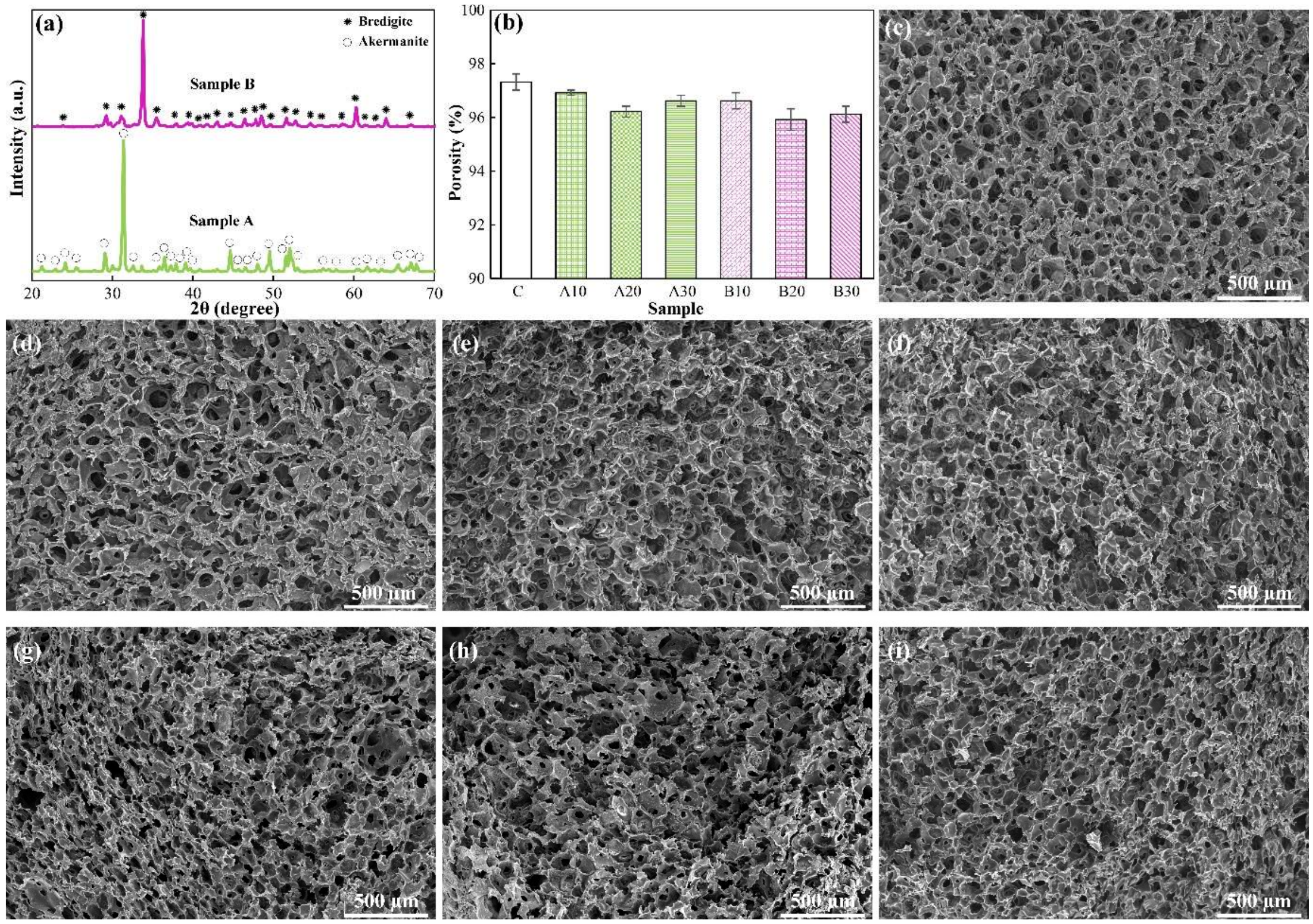

*Fig. 1. XRD pattern of Powders A and B (a). Archimedes' porosity of the scaffolds (b). SEM micrographs of the manufactured scaffolds: C (c), A10 (d), A20 (e), A30 (f), B10 (g), B20 (h), and B30 (i).*

Fig. 1b also shows the porosity of the gelatin-based scaffolds, measured by the Archimedes' immersion method. It is evident from the data that the samples exhibit a porosity within the range of almost 96-97%, with an insignificant decline with the addition of the silicate powders. The SEM images of the samples (Fig. 1c-i) reveal intricate textures of interconnected pores uniformly distributed across the surfaces, resembling a delicate mesh. The micrographs also showcase pore sizes ranging in 50-300 µm, where the addition of the silicate powders slightly reduces the pore size. An optimal freeze-drying process forms such pore architectures by freezing the solution and then sublimating ice crystals under controlled reduced pressure and temperature. The incorporation of ceramic particles reduces the chance for the formation of large pores via occupying interstitial spaces within the gelatin matrix and acting as nucleation sites for ice crystals [48, 49]. Overall, such pore networks are considered promising for tissue engineering in both morphology and dimension, as they can effectively facilitate cell infiltration, support nutrients and oxygen transport, enhance waste removal, mimic natural tissue structures, and promote seamless integration with native tissues [50-52].

### *3.2. Physiochemical Analyses*

The hydrolytic degradation behavior of the samples in PBS is demonstrated in Fig. 2a. As observed, the introduction of the silicate powders regulates the degradation level of the gelatin-based scaffolds, with this reduction being more pronounced when employing akermanite. Processing of gelatin with EDC and NHS results in the formation of amide bonds between gelatin molecules, establishing a cross-linked network with increased integrity [53, 54]. However, the attained enhanced stability is still inferior to that of silicate bioceramics like akermanite and bredigite. Additionally, owing to lower hydrophilicity and reduced water absorption capability, these ceramics act as a physical barrier and restrict water access to gelatin molecules. The stoichiometry of these two silicates distinctly reveals the higher level of silicon and magnesium but the lower level of calcium in akermanite compared to bredigite. Given the higher covalent character of Si-O and Mg-O bonds than Ca-O bonds [55, 56], the higher stability of akermanite is inferred, explaining its stronger impact on modulating the degradation of gelatin. According to the literature [57-59], the biodegradation rate of the composite samples falls within an optimal range for effectively supporting cellular responses, tissue regeneration, and integration in the context of tissue engineering.

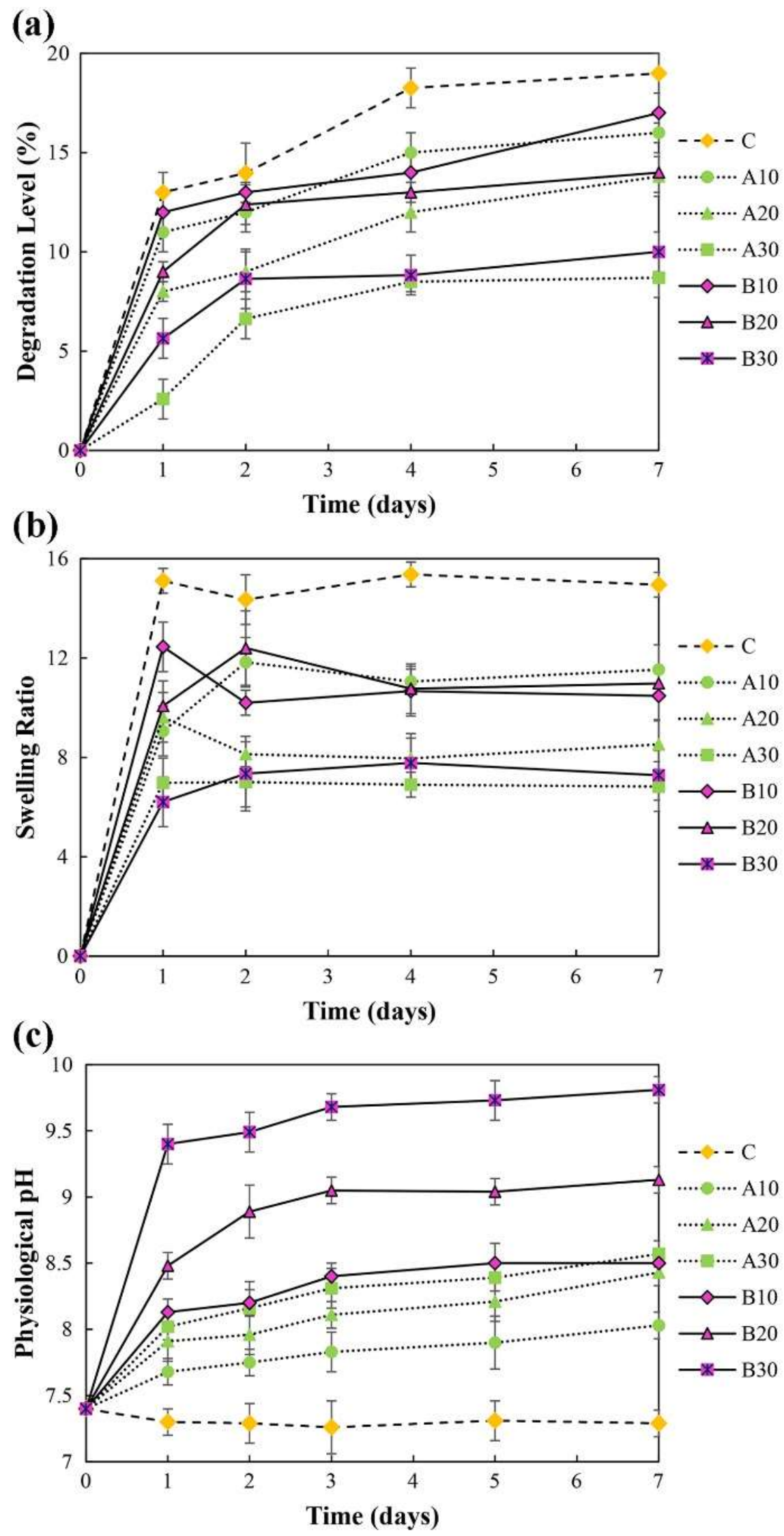

*Fig. 2. Degradation (a), swelling (b), and physiological pH variation (c) profiles of the different scaffolds.*

As well as lower hydrophilicity and reduced water absorption ability, the ceramics lack the swelling and hydrogel forming abilities in comparison to gelatin. Consequently, the incorporation of these ceramics results in a decreased swelling degree of gelatin, as depicted in Fig. 2b. The decline in the swelling rate at high exposure times is attributed to the dominance of degradation leading to a reduction in the measured weights ($W_w$), a time-dependent factor integrated into Eq. 2. The observed levels of swelling in the gelatin-based scaffolds are promising for tissue engineering, due to facilitating the formation of hydrogels and allowing them to mimic the extracellular matrix for supporting critical processes like cell attachment, adhesion, proliferation, and tissue regeneration [60-63].

Fluctuations in physiological pH, influenced by the degradation rate of biomaterials and resultant byproducts, play a critical role in governing cell viability and functionality, enzymatic activity, inflammatory responses, tissue integration, and overall biocompatibility. Controlling and maintaining an environment close to physiological pH is essential for optimizing the performance of biomaterials, minimizing adverse effects, and ensuring its compatibility with biological systems. Fig. 2c presents variations in the pH of PBS in contact with the scaffolds. A slight decrease in pH is observed for the gelatin samples due to the hydrolytic cleavage of peptide bonds, leading to the release of amino acid molecules. Nevertheless, the incorporation of the Mg-Ca silicates tends to elevate pH toward basic values. This pH increase is particularly pronounced when utilizing bredigite, as it releases higher quantities of alkaline calcium ions into the medium [64-66].

### *3.3. Cytocompatibility Assay*

Fig. 3a depicts the MTT results of the scaffolds. It is observed that all the samples demonstrate HUVEC metabolic activity comparable to the control (C), typically exceeding 85%, throughout the entire duration of cell culture. This suggests the promising potential of the scaffolds in fostering a conducive microenvironment for HUVECs, which is crucial for vascularization. Based on ANOVA, on the first day of culture, Samples A30 and B10 provide higher metallic activity for HUVECs than the other scaffolds, while the remaining samples exhibit no significant differences among each other. This meaningful superiority is observed for Samples A10 and A20 on the second day. On the third day, Samples A10 and A20 further enhance their superiority, while the cell metabolic activity for Samples A30, B10, B20, and B30 is considerably reduced. Accordingly, it is concluded that Samples A20 and A30 offer the most substantial enhancement in the metabolic activity of HUVECs.

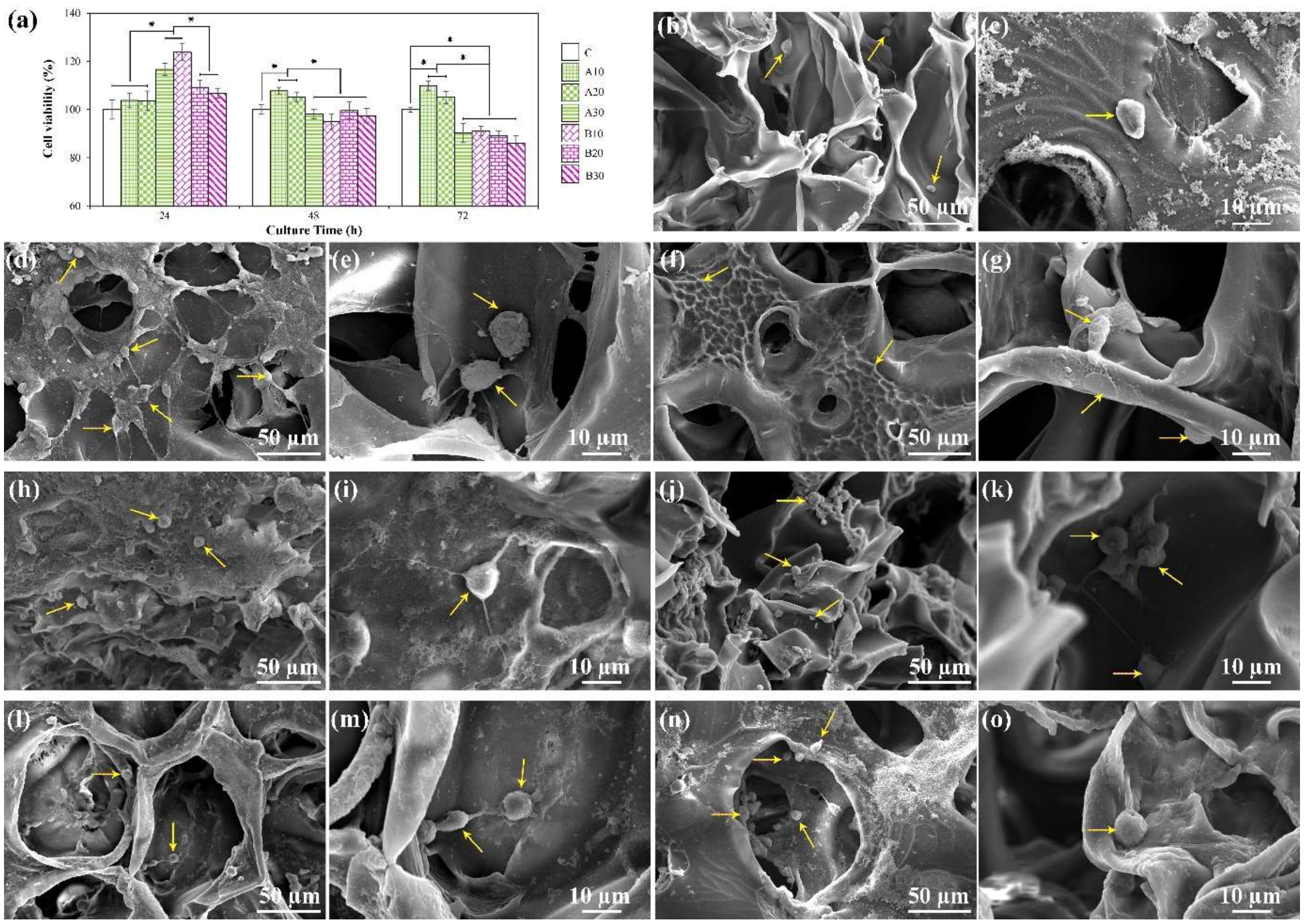

*Fig. 3. MTT assay results for the scaffolds with * denoting statistically significant differences (a). SEM morphology of HUVECs cultured on the samples: C (b, c), A10 (d, e), A20 (f, g), A30 (h, i), B10 (j, k), B20 (l, m), and B30 (n, o) in two magnifications, with arrows indicating typical cells.*

To elucidate the observed trend, several promoting and deteriorating factors in the cell viability and metabolic activity of HUVECs should be considered. According to the literature, the constitute ions of the used bioceramics, including Mg [67, 68], Ca [69, 70], Si [71, 72], and Cu [37, 39] have positive effects on the activity of HUVECs, with the threshold value of the latter being critically low before reaching a cytotoxicity level [73, 74]. Also, the alkalization of culture media due to the degradation of exposed biomaterials, including Ca-Mg silicates [64-66], has been shown to diminish biocompatibility. Taking into account these promoting and inhibitory factors, it appears that on the first day, the release amount of the beneficial ions from Sample A10 and A20 is not sufficient to significantly enhance cellular activity, as the biodegradability of akermanite is lower than that of bredigite. However, the high content of Cu-doped akermanite in Sample A30 compensates for its relatively slow degradability, contributing to a noticeable positive effect. Given the higher biodegradability of bredigite, Sample B10 depicts a favorable behavior, but the excess alkalization of Samples B20 and B30 (Fig. 2c) and/or the excess release of copper can be responsible for the reduced cell cytocompatibility. The inhibitory influence of bredigite on the viability of other cell types, including human bone marrow mesenchymal stem cells [75], dental pulp stem cells [76], and osteoblast-like cells [77], has been documented in prior studies. On the second day, the deleterious impacts of alkalization and/or copper release for Samples A30, B10, B20, and B30 prevail over the enhancing role of Ca, Mg, and Si ions, while the release amount of the beneficial ions for Samples A10 and A20 reaches sufficient levels. The same trend continues and is intensified on the third day of cell culture.

### *3.4. Vascularity Assessments*

HUVEC attachment is a foundational step for the initiation of vasculogenesis and angiogenesis in scaffold-based tissue engineering [78, 79]. The SEM micrographs of HUVECs cultured and fixed on the scaffolds are indicated in Fig. 3b-o, including two magnifications for each sample. Upon first observation, all the specimens represent several cells attached to their surfaces due to their cell cytocompatibility (Fig. 3a), albeit with different morphologies. On the gelatin sample, scattered cells with no evident cellular protrusions are observed (Fig. 3b), which is not regarded as a favorable cell adhesion. Also, the presence of debris near cells (Fig. 3c) confirms this assignment, which can be due to cell damage, death, or disintegration. Given the excellent biocompatibility of gelatin, this suboptimal feature can be ascribed to the presence of residual chemical cross-linking agents [80, 81].

Upon introducing the ceramic powders, the appearance of filopodia/lamellipodia around cells and their increased proximity imply enhanced interactions between the cells and surfaces owing to the beneficial influences of Mg [67, 68], Ca [69, 70], Si [71, 72], and Cu [37, 39] in terms of supporting cellular metabolism, regulating signaling pathways, stimulating extracellular matrix production, exhibiting antioxidant effects, and enhancing cellular adhesion and proliferation. The characteristic pattern of well-organized actin filaments is more pronounced in Samples A10 and A30, as depicted in Figs. 3d and 3h, respectively. The most distinct morphological feature is evident in Sample A20 (Fig. 3f), showcasing a so-called HUVEC cobblestone phenotype. This *in-vitro* closely packed monolayer arrangement of cells suggests their robust health, migration, signaling, and well-differentiated state as they mimic an endothelium *in-vivo* [82-84], accompanied by a widely spread-out cell morphology on narrower struts (Fig. 3g). The further enhanced cell adhesion observed in the akermanite-containing groups, in comparison to the bredigite-containing ones, provides additional confirmation for the MTT cell cytocompatibility assay (Fig. 3a).

*In-vitro* capillary tube formation by HUVECs is a commonly used experimental model that mimics the early stages of blood vessel formation *in-vivo*. Under favorable conditions, cells attach within the first hour, initiate migration toward each other over the next 2–4 hours, and subsequently form capillary-like tubes maturing between 6-16 hours [85]. The illustrative images of the tube formation assay using the medica collected from the samples are shown in Fig. 4. In the negative control (Fig. 4a), a lack of tube formation is evident due to the inhibitory effect of DMSO that suppresses matrix metalloproteinase-2 production [86, 87]. In contrast, the positive control using VEGF displays a stimulated cellular organization resembling tubes (Fig. 4b), attributed to the activation of signaling pathways that drive vascularization [7, 8]. The akermanite-containing samples exhibit favorable tube formation with lumens appearing as interconnected networks, comparable to that of the positive control and superior to the bredigite-containing scaffolds. Typically, Samples A10 and A20 exhibit the highest stimulating effect, characterized by an increased number of branching points and tubes, greater complexity and interconnectivity of the formed tubular-like networks, and thicker tube walls. As can be seen, the ranking of the samples in promoting tube formation is in good agreement with the results observed in the MTT and cell attachment assays (Fig. 3).

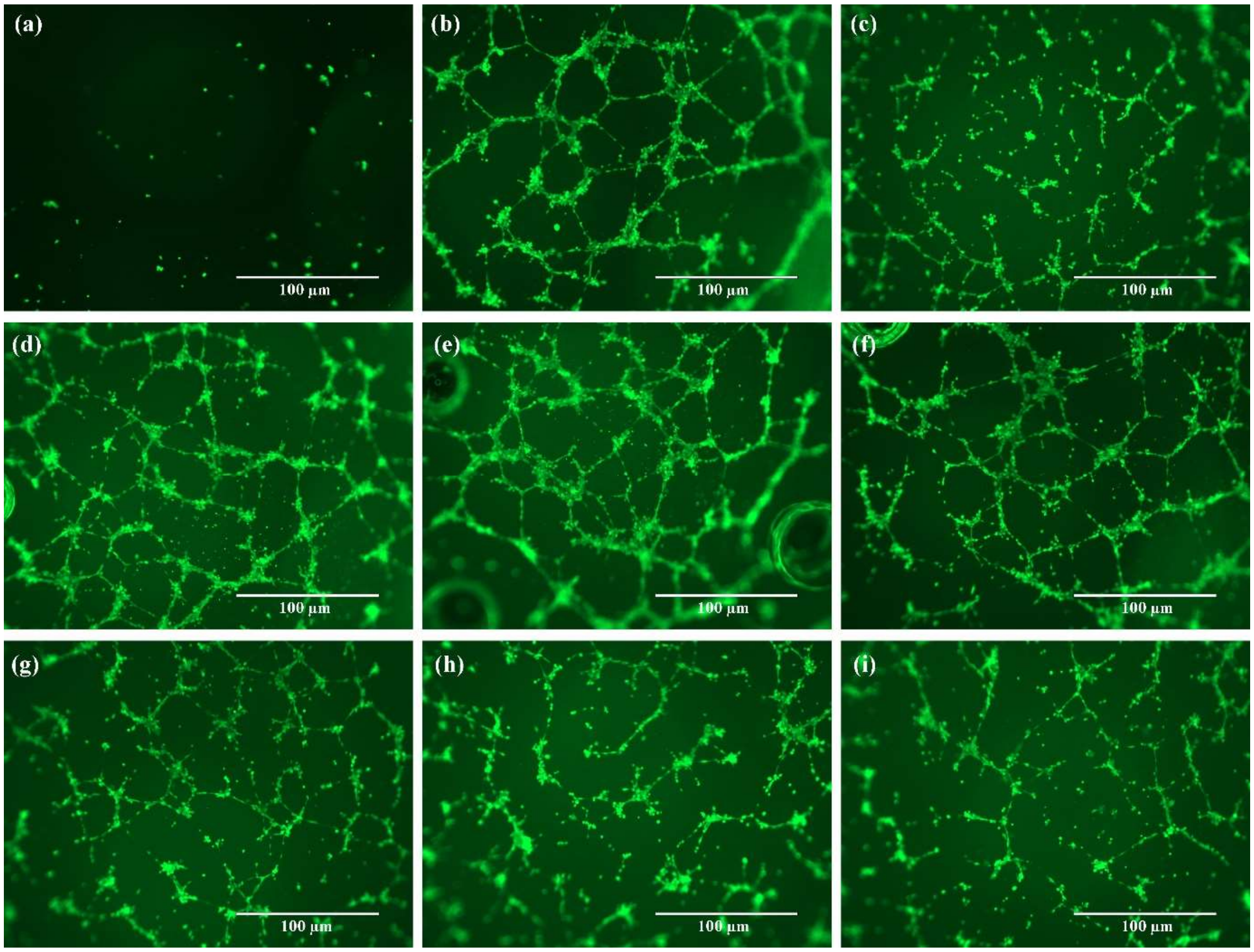

*Fig. 4. Behavior of HUVECs plated on the Matrigel matrix with the extracted media: the negative control (a), positive control (b), C (c), A10 (d), A20 (e), A30 (f), B10 (g), B20 (h), and B30 (i).*

Fig. 5 represents the histological development of angiogenesis in the chick embryos exposed to the samples over a period of 0 to 48 h. The analysis reveals distinct hierarchical variations among the samples based on the density, branching, complexity, length, diameter, and capillary formation of vessels. Notably, Samples A10 and A20 exhibit the highest levels of vascular network development, followed by A30. Sample C holds the subsequent position, indicating a significant enhancement in *ex-ovo* vascularity of gelatin with the addition of Cu-doped akermanite. In contrast, Samples B10, B20, and B30 not only fail to develop vascularization but also destruct the pre-existing networks. As observed, the hierarchical ranking of the samples in *ex-ovo* chick embryo angiogenesis aligns thoroughly with the outcomes of the MTT cytocompatibility (Fig. 3a), cell attachment (Fig. 3b-o), and tube formation (Fig. 4) assays.

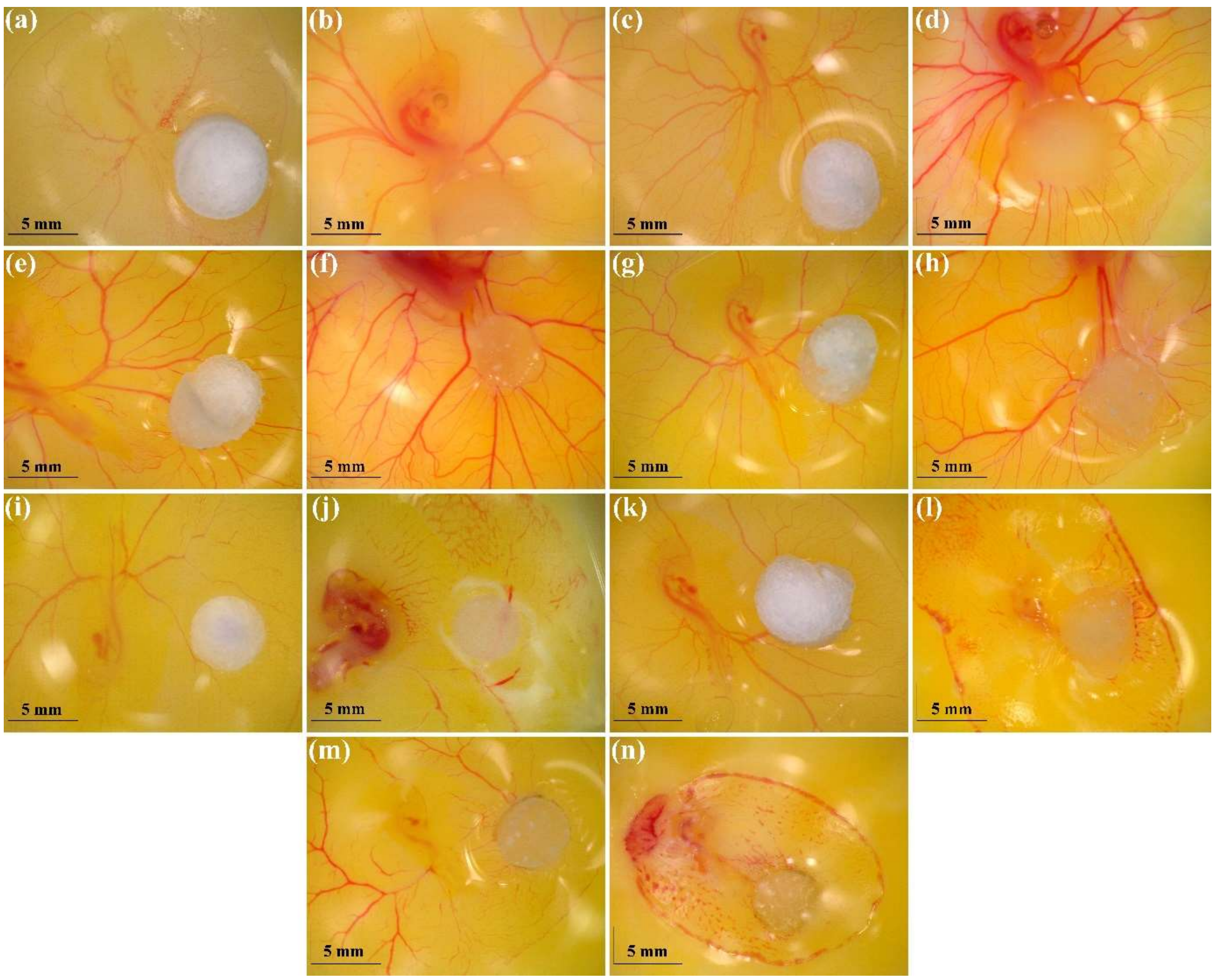

*Fig. 5. Dynamic snapshots of ex-ovo cultured chick embryonic development under the influence of distinct samples: C (a, b), A10 (c, d), A20 (e, f), A30 (g, h), B10 (i, j), B20 (k, l), and B30 (m, n) at 0 and 48 h for each group, respectively.*

## 4. Conclusion

This study explored the potential of enhancing vascularization in bone tissue engineering scaffolds by incorporating akermanite and bredigite, both doped with 5% Cu, into gelatin.

Physicochemical analyses indicated that the addition of the silicate powders regulated gelatin degradation and swelling, while also increasing the pH of the surrounding medium. Cytocompatibility assays underscored the enhanced metabolic activity of HUVECs on the scaffolds containing 10 and 20% Cu-doped akermanite, where favorable cell adhesion and morphology were observed. Assessments of vascularity showed improved HUVEC attachment, tube formation, and vascularization in *ex-ovo* chick embryo chorioallantoic membrane assays on the scaffolds with Cu-doped akermanite, especially at 10 and 20% incorporation levels. These results indicate that Cu-doped akermanite is promising for addressing vascularization challenges in tissue-engineering scaffolds by providing an optimized release of pro-angiogenic ions. Future research should focus on examining the long-term effects and *in-vivo* performance of these scaffolds to confirm their effectiveness in clinical settings.

## Acknowledgments

L.T. acknowledges the support from National Institutes of Health under award numbers R56DE029191 and R21EY035480.